*Review*

# Energy Efficient AI-Enabled Wireless Sensor Networks for Mission Critical Environments: A Systematic Review across Smart Grid, AI, and Urban Infrastructure Applications

**Alexandros Gazis [1,*], Valeri Mladenov [2] , Kleanthi Santamouri [3], Stylianos Pappas [4,*]**

[1] Department of Electrical and Computer Engineering, Democritus University of Thrace, Greece ; agazis@teemail.gr
[2] Department Fundamentals of Electrical Engineering, Technical University of Sofia, Bulgaria; valerim@tu-sofia.bg
[3] Salvezza Energy Systems LTD, Pahpos, Cyprus; a.santamouri@salvezza.eu
[4] School of Engineering, Merchant Marine Academy of Aspropyrgos, Greece; s.pappas@aenynanp.gr
* Correspondence: agazis@teemail.gr, A.G, s.pappas@aenynanp.gr, S.P

**Abstract**

Advanced wireless sensor networks powered by artificial intelligence are increasingly required for applications demanding continuous monitoring, autonomous operation, reliable communication, and fast decision support. This systematic review examines recent work from 2023 to 2026 on energy-efficient, AI-enabled wireless sensor networks (WSNs) in mission-critical environments, with particular focus on power electronics, and urban infrastructure systems. The authors synthesise a corpus of 50 DOI indexed studies satisfying inclusion criteria that received qualitative thematic coding and comparative analysis. Other references were only cited to provide historical, methodological, or technical context and were not included in the systematic review corpus. As such, our results show that AI can improve WSN energy behaviour through routing and clustering, edge AI, reinforcement learning, fuzzy logic, metaheuristic optimisation, and AI-based security. At the same time, energy efficiency cannot be treated as an isolated performance target. In mission-critical systems, security, latency, and reliability are closely interlinked requirements. The review concludes that future work should move away from optimising protocols in isolation, and instead focus on building lightweight, explainable, secure, and field-tested AI-driven WSN architectures suited to real operational environments.

## 1. Introduction

Wireless sensor networks are used to collect, transmit, and process data in applications that require real-time monitoring and control. At first glance, a WSN may appear to be a simple arrangement of sensor nodes, communication links, and a sink or gateway that forwards the collected data to an application [1]. However, when such a network operates in a mission-critical environment, its role becomes considerably more demanding. A failure may result not only in the loss of measurements, but also in reduced situational awareness, interruption of critical services, safety risks, or delayed decision-making. For this reason, WSNs are increasingly used as distributed sensing infrastructures that support real-time and near-real-time decisions in surveillance systems [2,3], smart grids [4], and urban infrastructure applications [5].

Energy remains one of the principal technical constraints of WSNs [6], as sensor nodes typically have limited processing capacity, memory, radio resources, and battery power. This limitation becomes especially important when nodes are deployed in locations where maintenance or replacement is costly, dangerous, or impractical [7]. Such environments include remote sensing areas [8], power-line monitoring systems [9], underground [10] and outdoor urban installations [11], industrial facilities, and disaster-affected regions. Consequently, extending network lifetime requires not only reducing overall energy consumption but also preventing premature node failure and balancing energy use across the network [12,13]. Nevertheless, low energy consumption alone is insufficient in mission-critical applications, where packet loss, delayed responses, or security vulnerabilities may directly affect operational continuity.

Artificial intelligence provides a more adaptive means of addressing these interconnected requirements [14]. Machine learning, reinforcement learning, fuzzy logic, metaheuristic optimisation, neural networks, graph-based models, and edge AI can support routing, cluster-head selection, resource allocation, anomaly detection, and adaptive network control [12,15–17]. Rather than relying exclusively on static decision rules, an AI-enabled WSN can adapt its operation according to residual energy, link quality, congestion, node density, mobility, security conditions, and application requirements [18,19]. Energy efficiency should therefore be understood as a system-level coordination problem involving sensing [20], computation [21], communication [22], and security [23], rather than merely as a reduction in radio use.

The application domains evaluated in this study manifest these requirements in various ways. Wireless Sensor Networks (WSNs) have applications in intrusion detection, monitoring of environment, tracking of assets and gathering of tactical data, in surveillance and monitoring scenarios where operation must also be energy efficient [24]. In smart grids, WSNs have been adapted for PLM, smart metering, and cyber-physical security [25]; consequently, attacks such as eavesdropping [26], selective forwarding [27], or data falsification [28] may result in immediate operational impact. Some applications for urban infrastructure include pollution monitoring, traffic management, smart lighting, waste management, early-warning services and public safety [29–31], where large-scale deployment also requires low cost, scalability and long-term energy efficiency [32,33]. An introduction of these domains is provided here to demarcate the scope of the review; a further detailing is provided in the coming sections.

Sensing that’s critical to the mission continues to get extended to heterogeneous and highly mobile communicative environments from conventional terrestrial WSNs. For example, in hybrid satellite networks, sensing information from different domains can be aggregated through cross-domain data aggregation, while dynamic group-key agreement enables confidentiality and secure changes in group membership. Similarly, authentication and rapid trust establishment in UAV-based flying ad hoc networks should be lightweight. Because their topology, connectivity, and relations vary continuously. As a result, we may argue that these examples show that secure aggregation, authentication, latency, reliability, and resilience cannot be designed independently from energy-efficient communication. This is important as, based on the above, it is evident that HCI use has a much broader relevance for mobile applications as it provides a unique set of challenges which must also be considered in the design phase.

While recent literature is abundant, it has also become fragmented. While some studies refer routing protocols, others may refer clustering, security, edge processing, or application monitoring [34,35]. Also, evaluation criteria vary to a large extent. A paper may target network lifetime [36]; another may target latency packet delivery ratio [37], attack detection accuracy [38] or coverage [39]. It creates challenges for direct comparison. It also creates the need for a review not only of algorithms, but also of the performance-fitting of energy-efficient AI-enabled WSNs in various mission – critical environments.

As such, the objective of this study is to explore some of the recent research activities on energy-efficient AI-enabled WSNs in energy, smart grid and urban infrastructure applications. This research explores the main technological approaches, contrasts the needs of the three application sectors, and identifies recurring trade-offs and existing gaps of research. The issue of how the deployment of AI can contribute to energy performance without threatening a system's security, reliability and operational readiness will be examined. Most notably, this issue is of great importance as most critical systems in real life are not optimized for one metric. On the contrary, they should continue to operate despite limitations, defects, and occasionally adversarial situations.

This review is thus centered not on how artificial intelligence or wireless sensor networks in isolation, but focuses on AI-enabled WSNs deployed for mission-critical environments. As such, the review examines routing, clustering, edge processing, security, and resource allocation based on their capabilities of supporting energy-efficient, reliable, secure and timely operation under critical conditions. Lastly, surveillance and monitoring systems, smart grids, and urban infrastructure are regarded as distinct application domains through which this central problem is comparatively assessed, rather than separate or parallel research topics.

The rest of the paper is structured as follows: Section 2 presents a literature survey mapping the key technical trends in AI-enabled WSNs, covering routing, clustering, edge AI, security, and related areas. Section 3 defines the research questions and the scope of the review in terms of time period, technologies, and application areas. Section 4 describes the methods used for source selection, assessment, extraction, and synthesis of information. Section 5 presents the findings through thematic coding and comparative tables. Section 6 discusses the principal technical and operational trade-offs introduced by the use of AI in mission-critical WSNs. Finally, Section 7 concludes the paper and outlines directions for future research.

## 2. Literature Review

Recent research on energy-efficient AI-enabled WSNs builds on four overlapping strands: routing optimisation, energy-aware clustering, edge-based intelligence, and security in critical or semi-critical environments. Most real systems exhibit overlap between these constructs. For example, a routing decision [40] may influence network energy balance, latency, and the exposure of data to untrusted nodes. Likewise, a security solution [41] may improve robustness but introduce additional processing and communication costs. While energy efficiency [42,43], is often framed as a low-power hardware problem, the literature from 2023 to 2026 shows the opposite to be true: it is increasingly treated as an optimisation issue at the network and system level.

Routing is the most mature area of the literature [44,45]. The reason is straightforward as radio communication is typically the costliest operation for a sensor node. If packets are routed along poor available paths, or if the same nodes are repeatedly overused as relays, the network can lose coverage even while energy remains available elsewhere in the system. The authors in [12], reviewed AI-driven, energy-efficient routing approaches in IoT-based WSNs, where routing decisions are made based on remaining energy, link quality, distance, traffic load, and network density. The study [13], shows in their study of AI-based cluster routing that fuzzy heuristics, metaheuristics, and machine learning models can limit uneven energy consumption. This is especially relevant in critical systems, where the failure of even two nodes or more may create a blind spot or disconnect too early.

As such, recent research from 2026 reinforces the above-mentioned trend. The study [46] proposed an adaptive routing strategy using more than 1 algorithm combining clustering and routing within a PSO-based, fuzzy, self-organising, energy-efficient framework [47]. The study, [48] proposed a router design approach based on chaotic and bio-inspired principles for energy-efficient routing in wireless body sensor networks (WBSNs). While a body sensor network is not equivalent to all available WSN infrastructures, such as a sensor network monitoring building temperature, both share a key design constraint: reliable sensing under very limited energy budgets.

Reinforcement learning is also growing in prominence, valued for its ability to let networks adapt future decisions based on past states. The authors in [49] apply reinforcement learning to WSN routing, while [50] proposes a multi-agent reinforcement learning approach to energy-efficient WSN routing. The benefit of these approaches is adaptability, but training overhead, reward design, and execution cost remain non-trivial limitations. An algorithm that performs well in simulation may not be practical to implement if learning is too expensive or too slow in low-power networks.

Another important research stream is edge AI [51]. Edge-enabled WSNs process part of the data at the source rather than sending all raw data to the cloud. According to [16], statistical methods, edge AI and 5G/6G smart sensor networks are integrated. According to [29], a smart city architecture can leverage edge AI for locally organized event processing, which helps reduce latency, bandwidth consumption, and privacy risk. In a WSN, a gateway, cluster head, or edge node can filter data, detect anomalies, and forward only the useful information. This can reduce communication energy, though part of that cost shifts to computation instead. As a result, where AI capability is placed within the network becomes a critical design choice.

The practical application of ideas is through urban infrastructure. The network of sensors presented by [30] is a low-cost AI-IoT pollution-monitoring sensor that supports citizens with respiratory problems. The study, [31] reviews WSNs for urbanization and discuss traffic management, disaster management and surveillance applications. According to a study, the green and blue LEDs are regarded as the most effective in reducing light pollution. The communication networks that WSNs can offer complement the benefits of other public service systems. As a result, energy efficiency has social, economic and service consequences.

Smart grids are a more security sensitive environment. A recent study, [52] proposes a GRNN-based detection of eavesdropping attacks in smart grid WSNs [53] examines cybersecurity challenges in smart grids based on WSNs. According to [54], the integration of smart grid and green hydrogen enabled through AI. Ultimately, sensing data plays an increasing role in system level energy control. The authors in [55], suggested a framework for smart grids that is aware of energy consumption and based on the Internet of Things. The network topology of WSNs correlates with the stability of power systems in smart grids. Efficient energy use is inextricably linked to secure and reliable data flows.

Work on WSN oriented to security is also important for many mission-critical cases. For example, the authors in [56], focused at selective forwarding attack through attention-based detection. In addition, sensor networks have been covered for secure optimal routing and trust-based routing [57,58]. In tactical or adversarial environments, energy-saving routes are useful only if also trustworthy. Accordingly, future WSNs should not optimize energy first and add security later. The requirements for the system need to be designed together.

In general, there was progress although uneven maturity. Both routing and clustering are well-developed; however, real-world evaluation [59], common benchmark programs [60] and energy-security [61] joint assessment are lacking. Many works still focus mostly on simulations with varying assumptions that are hard to compare. The literature is treated thematically and comparatively, rather than as a mere list of proposed protocols.

In addition to individual routing techniques, recent literature also points to a broader shift from classical WSNs to multi-layer AIoT architectures [62], whereby energy efficiency increasingly depends on how tasks are distributed. In the conventional design of WSN, it was generally assumed that the sensor node collected data and routed it toward the sink or gateway. Most of contemporary AI enabled designs have more decision points. The sensor (e.g. camera) may do light filtering, the cluster head may aggregate data, the edge may run anomaly detection, and the cloud may do the heavy analytics or model training [16,29]. In mission-critical environments, distributed logic is particularly important [63], as it minimises reliance on a single central infrastructure and allows quicker reactivity when the network detects an important event.

According to the literature, energy optimization techniques should be seen as architectural choices rather than simply as algorithms. Reinforcement learning can be useful where there are sufficient state in-formation and a stable system reward design but may not be suitable for a highly constrained or unstable network [49,50]. Graph neural networks can also account for relationships between the nodes inherently; however, they require careful attention to computational budgets [64]. According to [52,53], [65], reliability and security of measurements are generally the first priorities in smart grids, while in the case of urban infrastructure, the most important issues are [30,31], [66] scalability, cost, and long-term maintenance. In surveillance or monitoring type cases, energy efficiency is linked to stealthy, autonomous, resilient operation in environments where nodes cannot have repairs or replacements done easily [57,58], [67].

Thus, the review needs to compare not only the techniques themselves, but also the level of the architecture at which each technique is applied and the type of benefit it provides. Fewer transmissions, lower computation, superior path selection, fewer retransmissions, data aggregation, and premature filtering of unimportant data may save energy. Every mechanism has a different cost and tells a different operational story. An edge-based model is potentially a good fit for smart city monitoring but probably not for small tactical sensor nodes with very low power. In a monitoring application setting or smart grid, a secure routing protocol may be warranted, but it may be overkill for simple environmental sensing. Table 1 helps in understanding these architectural nuances.

**Table 1.** Comparison of architectural layers and AI-enabled energy optimization in WSNs.

| Architectural layer | Main AI function | Energy benefit | Best-fit domain | Main risk |
|---|---|---|---|---|
| Sensor node | Local filtering, lightweight inference, periodic sampling | Reduces unnecessary measurements and transmissions | Urban sensing, tactical monitoring | Limited processing capacity |
| Cluster head | Node selection, aggregation, energy balancing | Reduces long-range transmissions to sink/gateway | Smart grid, static WSNs | Cluster reformation overhead |
| Edge gateway | Anomaly detection, event classification, local decision-making | Reduces cloud traffic and latency | Smart grid, urban infrastructure | Shifts cost to edge devices |
| Network/control plane | Adaptive routing, resource allocation, security-aware policies | Improves path and resource use | Mission-critical deployments | Higher complexity and need for reliable state information |

Table 1 exhibits an indicative correspondence to the OSI communication model of the architectural layers. Specifically, the functions of a sensor-node mainly from the physical and data-link layers that include sensing, radio transmission, medium access, and local connectivity though local filtering and lightweight inference may introduce processing at the application-layer. This means that the main function of operations that needs to be operated is in the data-link and network layers, cluster heads coordinate access, aggregate traffic, choose routes, and forward data. More specifically, the edge gateways are the devices which spread in network, transport and application layers since they connect WSN traffic with external networks, as well as support anomaly detection, event classification and local decision making. As such, the functions associated with the network and control-plane concern the data-link and network layers to a significant extent. Lastly, adaptive routing, resource allocation, topology management as well as security-aware policies come into play. As a result, because AI-enabled WSN architectures (generically) allocate functionalities in multiple OSI layers rather than allocating each in one OSI layer, this correspondence should be characterized as approximate.

Our review draws on a small set of representative cases for clearer comparison across applications. The literature discussing secure routing, tactical monitoring, extreme environments, and industrial cyber-physical systems ostensibly refers to monitoring applications, while studies of cybernetics smart grid and urban infra-structure tend to be more explicitly labelled. Table 2 helps in understanding the three indicative cases served as interesting analytical examples to compare with relevant research work and not as new empirical experiments. In every case, we identify the corresponding technology domain and evaluate its relevance to smart-grid and urban-infrastructure use cases. Analytically, the routing, sensing, edge-computing, security and operation requirements of the different applications help to substantiate and discuss their energy-related characteristics. As such, the cases are not used for new empirical studies but for comparison.

**Table 2.** Indicative case studies and their alignment with the three review domains.

| Case study / application | Technology domain | Smart Grid | Urban Infrastructures |
|---|---|---|---|
| Tactical perimeter and asset monitoring with energy-aware secure routing | Secure WSN, trust-aware routing, edge alerts | Indirect relevance: protection of critical nodes and infrastructure | Moderate relevance: public safety and emergency monitoring |

| Power transmission line and smart grid monitoring with edge computing | Smart grid WSN, SWIPT/edge sensing, anomaly detection | Direct relevance: real-time monitoring, reliability, cybersecurity | High relevance: continuity of urban energy services |
|---|---|---|---|
| Urban pollution and adaptive lighting sensing | AI-IoT, low-cost sensors, smart city edge AI | Indirect relevance: energy management of municipal services | Direct relevance: scalability, cost, sustainability |

## 3. Research Question and Scope

The main research problem around which this review is organized is that AI-enabled WSNs must achieve high energy efficiency without sacrificing reliability, security, low latency, and operational continuity in mission-critical environments. This question matters because WSNs are increasingly used in applications well beyond simple data collection. More and more, real-time decisions in smart grids, monitoring and surveillance, and urban infrastructure depend on sensor data, meaning the quality of network performance directly affects the quality of decision-making in these contexts.

The research question has three closely associated levels. The first refers to the level of technique that refers to artificial intelligence algorithms, networking architecture, routing, clustering, and sensor operation. The second, is the level of application takes into account the operational requirements of monitoring systems, smart grid, city infrastructure, etc. The third level refers to the evaluation level, which looks at how performance is evaluated through energy efficiency, reliability, latency, security, coverage, etc.

As such, the main research question is addressed through five sub-questions:

1. Which AI-enhanced techniques are used for energy optimisation in WSNs? The current literature identifies several relevant techniques, including machine learning, reinforcement learning, fuzzy logic, neural networks, metaheuristic optimistion, graph neural networks, and hybrid clustering-routing approaches [12,13], [46,47], [64]. These techniques support routing, cluster-head selection, resource allocation, energy consumption forecasting, and anomaly detection. The central concern is not which approach performs best, but where it fits within the architecture and which application it suits.
2. What about concerns in regards to energy metrics? Common metrics adopted in WSN research include total energy consumption, network lifetime, time to first node death, residual energy, number of alive nodes, energy per packet, load balance, and packet delivery ratio [12,13], [68]. However, in mission-critical use cases, measures beyond energy — such as delay, reliability, coverage, resilience, and security — are equally important. A protocol that lowers energy use by cutting back transmissions is unacceptable if doing so delays alarms or compromises data integrity. This review therefore treats energy efficiency as a system property rather than a single number.
3. What are the differences between our article's application domains? In monitoring and surveillance environments, WSNs must operate under uncertain conditions in mobile and hostile settings, where secure routing, trust-aware routing, and attack detection are essential, since routes must be both energy-efficient and secure. A key requirement for smart grids is continuous monitoring of generation and consumption, alongside technical accuracy, cyber-physical stability, and resilience against attack [52-55], [65]. Urban infrastructure is constrained mainly by scale, cost, interoperability, dense coverage, and long-term maintainability [29-31], [66]. The same underlying energy problem is therefore prioritized differently across domains.

4. What about recent trends such as AI and case-specific edge AI? Performing computation close to the data source reduces retransmission and latency while improving privacy [16,29]. However, edge AI introduces a practical trade-off: reducing communication cost increases computational load at gateways, cluster heads, or capable sensor nodes. This raises the question of whether AI genuinely reduces total energy cost, or simply shifts the burden from communication to computation — a question of particular relevance for low-power deployments.
5. What are the research gaps to investigate? Many studies are assessed through simulation rather than field deployment. Numerous papers report energy savings without testing performance under attack, jamming, mobility, or hardware limitations. No single benchmark applies consistently across monitoring, smart grid, and urban infrastructure scenarios. As a result, the review employs a comparative framework that organises the literature by technique, energy contribution, application domain, metrics, and limitation.

As such, three criterions define the scope of the review. First of all, the time frame mentioned is 2023–2026 so the paper captures state-of-the-art work on AIoT, edge AI, smart grid, and intelligent sensor networks. In the second place, it is also the study of energy efficient, AI-based decision making, WSN sensor architecture, secure routing, and applications such as a smart grid, smart city, and urban solid waste management systems. Thirdly, the review also excludes general papers in AI which are unrelated to any sensor networks, communication, energy consumption or critical infrastructure.

The surveillance or monitoring component of the scope is interpreted in detail and in many parts but with a lot of caution. A lot of relevant literature does not use these terms strictly in titles. Rather, it focuses on secure routing, extreme environments, resilient WSNs, cyber-physical system, intrusion detection and disaster response. Even where these papers do not frame themselves around surveillance applications, the underlying concepts remain technically relevant to monitoring deployments, concerning autonomy, resilience, low maintenance, security, and operation under stress. The same logic extends to smart grids and urban infrastructure: the review includes studies that explicitly connect WSNs and AI to energy, monitoring, security, or operational decision support.

A broad scope makes the review more useful than a narrow protocol survey. The goal isn't to simply create a menu of algorithms but to provide the best AI-enabled WSN design for a mission-critical environment. Energy efficiency depends on the algorithm, architecture, and utilization of the hardware.

## 4. Methodology

This paper presents a comprehensive literature review focusing on energy-efficient AI-enabled wireless sensor networks for critical mission environments. The choice of these studies was made on account of the considerable difference with respect to the size of networks and simulation tools, energy models, performance metrics, and application assumptions. Combining direct numerical results could be deceiving. The review emphasizes the literature’s patterns, domains, trade-offs, and gaps, instead.

The review covers the years 2023–2026. This range was selected to capture initiatives and research building on developments from 2020 onward, spanning AI-enabled WSNs and intelligent urban sensing. Many earlier protocols are not well suited to machine learning, lightweight intelligence, or edge-based processing, though they remain historically relevant. A total of 50 studies met the predefined temporal, bibliographic, technological, energy-related, AI-related and application-domain criteria. Specifically, the formal coding of the studies facilitated their inclusion for the thematic synthesis, comparative tables and research gap identification. As such, the full reference list includes other publications that have been consulted to establish historical background, define well-established WSN concepts, outline methodological frameworks, e.g. PRISMA, JBI or provide contextual information. As a result, people, events and places referred to for background purposes and not examined formally are excluded from coding and comparative synthesis.

Simialriy, our search strategy targeted academic platforms and publishers in the fields of networking, sensors, AI, and cyber-physical systems. The two primary sources were IEEE Xplore and ScienceDirect, Elsevier and Springer. In addition, MDPI, arXiv, and TechRxiv were consulted, along with a smaller number of papers from Wiley Online Library, and ACM Digital Library. Moreover, preprint sources were treated with caution and included only where a DOI existed and the work was technically relevant to the topic - for example, reinforcement learning for energy-efficient WSN routing [50].

The search terms were organised into four clusters:

1. Cluster 1: covered network technology terms: "wireless sensor networks," "WSN," "IoT-based WSN," "sensor networks," and "edge sensor networks."
2. Cluster 2: covered energy-related terms: "energy efficient," "energy aware," "network lifetime," "low-power," "energy optimization," "energy consumption," and "green routing."
3. Cluster 3: covered AI techniques: "artificial intelligence," "machine learning," "deep learning," "reinforcement learning," "fuzzy logic," "metaheuristic optimization," "graph neural networks," and "edge AI."
4. Cluster 4: covered relevant application contexts: "mission critical," "surveillance," "smart grid," "urban infrastructure," "smart cities," "cyber-physical systems," "extreme environments," "secure routing," "disaster response," and "critical infrastructure."

After identifying a topic, the authors undertook an extensive literature search from the period 2020 to 2026 across IEEE Xplore, ScienceDirect, SpringerLink, MDPI, Wiley Online Library, ACM Digital Library, arXiv, and TechRxiv. The searched records were published from 2023 to 2026 in English language only. When supported by the database interface, the search was applied to the title, abstract, and author-keyword fields; otherwise, the equivalent query was applied to all indexed metadata. To signify the technological, energy-related, AI-related, and application-oriented aspects of the assessment, four Boolean query families were used. This was reflected in the query strings, search date, field restrictions, and number of records retrieved from each of the databases as shown in Table 3. Syntactic spelling specific to the relevant database was altered as needed, which did not alter the conceptual structure of the queries. The entire retrieved records were exported to the review register. Duplicates were removed, and remaining records were screened as per DOI, publication-period, thematic-relevance, and technical-adequacy criteria given below. Specifically, some of the exact Boolean query terms we used are the following:

- Artificial Intelligence-based Smart Energy-efficient Routing and Clustering

```
("wireless sensor network*" OR WSN OR "IoT-based WSN*" OR
"sensor network*" OR "edge sensor network*")
AND
("energy efficien*" OR "energy aware" OR "network lifetime" OR
"low-power" OR "energy optim*" OR "energy consumption" OR
"green routing")
```

```
AND
("artificial intelligence" OR "machine learning" OR "deep learning" OR
"reinforcement learning" OR "fuzzy logic" OR metaheuristic* OR
"graph neural network*" OR "edge AI")
AND
(routing OR clustering OR "resource allocation")
```

- Security and smart-grid WSNs.

```
("wireless sensor network*" OR WSN)
AND
("secure routing" OR eavesdropping OR "selective forwarding" OR
spoofing OR "attack detection" OR cybersecurity)
AND
("smart grid" OR "critical infrastructure" OR "mission critical")
AND
("energy efficien*" OR "energy aware" OR "network lifetime")
```

- Edge artificial intelligence and smart urban infrastructure.

```
("wireless sensor network*" OR WSN OR "IoT sensor network*")
AND
("edge AI" OR "edge computing" OR "local inference")
AND
("smart cit*" OR "urban infrastructure" OR pollution OR lighting OR
waste)
AND
("energy efficien*" OR latency OR scalability)
```

- Mission critical and extreme environments.

```
("wireless sensor network*" OR WSN OR IoT)
AND
("mission critical" OR surveillance OR monitoring OR
"extreme environment*" OR "industrial cyber-physical system*" OR
"disaster response")
AND
("green routing" OR "secure routing" OR resilience OR autonomy)
AND
("artificial intelligence" OR "machine learning" OR
optimisation OR optimization)
```

To extract information consistently, the review adapted a simple ETL logic to the technical literature review process. During the Extract stage, basic bibliographic metadata was recorded for each paper, including title, authors, year, publisher, DOI, application domain, and technical relevance. During the Transform stage, duplicate or unsuitable records were removed, and the remaining sources were classified according to AI technique, WSN function, energy metric, security relevance, and application domain. During the Load stage, the finalised records were captured into a consolidated reference register and used to build the synthesis tables and findings. This process does not substitute the technical judgement of the researcher, but minimizes in-consistency in selecting, evaluating and comparing sources. Our rationale and ETL steps are illustrated in Figure 1.

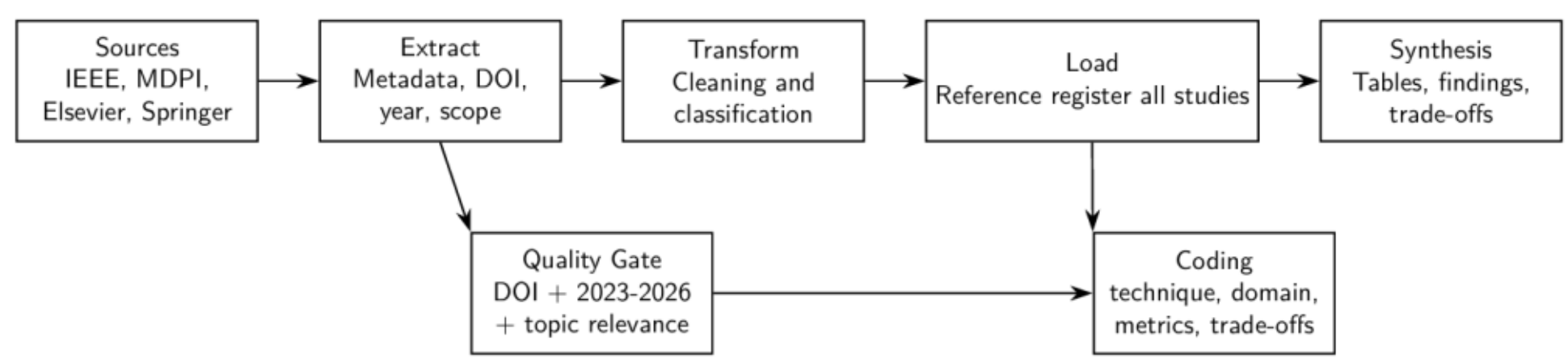

**Figure 1.** UML activity-style ETL logic for our review information and extraction.

Moreover, Data interpretation, claim assessment and the selection mechanism were organized in relation to the connection between search terms and rationale. In this sense, data interpretation refers to how a source was judged useful for the review: not merely because it mentioned WSNs, or AI, but only where it links a technical solution to both an energy dimension and application domain. The claim evaluation phase checked if each paper supported their claim with astute metrics like, network life time, residual energy, latency, packet delivery ratio, security detection accuracy etc. Only those sources that satisfied the DOI, temporal, thematic relevance, and technical adequacy were included in the Selection. As such, the search queries, rationale and most notably selection criteria are presented in Table 3.

**Table 3.** Search queries, rationale, selection criteria, and use in the review.

| Query / search direction | Rationale | Selection criterion | Use in the review |
|---|---|---|---|
| energy efficient AI wireless sensor networks routing clustering | Identify AI-based techniques for energy optimization | DOI, WSN/AI relevance, energy metrics | Literature review and results |
| secure routing wireless sensor networks smart grid attacks | Connect energy efficiency with security in critical infrastructures | Clear security/WSN link | Smart grid and monitoring-type discussion |
| edge AI smart city wireless sensor networks energy | Analyze local processing, latency, and scalability | Edge/urban infrastructure relevance | Urban infrastructure and ETL synthesis |
| mission critical IoT extreme environments green routing | Identify surveillance or monitoring-like technical environments | Resilience, autonomy, secure operation | Case studies and discussion |

The selection and presentation of sources followed a PRISMA-style logic, though the paper and it does not claim to be a meta-analysis. Specifically, the PRISMA flow diagram is used to show the sequence of identification, screening, eligibility, and inclusion. To ensure that evidence was analysed and synthesized in a transparent, documented and consistent manner, the JBI evidence-synthesis principles were used to guide this review paper. However, as the corpus comprised heterogeneous engineering, simulation-based, review and application-oriented studies, no one design-specific JBI checklist was applicable directly to all included publications. Consequently, we applied a technical quality-appraisal framework carefully throughout the review corpus [69,70].

Specifically, each study which potentially met the inclusion criteria was assessed with an adapted technical quality-appraisal framework developed for the heterogeneous WSN literature included in this review. More specific, the first criterion assessed is clarity in research objectives and technology contribution; sufficient description of proposed architecture/algorithm/dataset/experimental procedure; transparency in simulation/evaluation configuration (network size, energy model, parameters, or hardware when applicable; clear and suitable performance measure; consistency in methods, result, tables, figures, and conclusion; bibliographic/research integrity reliability (verifiable DOI, and not a retraction/expression of concern/duplicate publication/unresolved integrity matter.

Moreover, the scoring system for each criterion had the following conditions. Firstly, the sufficiently lacking characteristic got 0 value and so on. Similarly, the criteria deemed genuinely not applicable to the specific study design were designated as not applicable and excluded from the denominator. The maximum applicable score was used as the denominator to determine the normalised quality score. As such, any study that achieved a score of 75% or greater was considered high quality. Moderate quality was achieved by studies scoring between 50 and 74%. Low quality refers to any studies that achieved scores below 50%. The formal synthesis excluded studies of low quality, and studies were excluded if their bibliographic and research-integrity criterion score was 0.

Lastly, a definition of exclusion terms was also considered but was deemed as likely to reduce subjectivity. Analytically, a study was insufficiently clear if the text available did not permit identification of its objective of research, technical method, evaluation process, or main results. Moreover, a study was deemed internally inconsistent if there was an unresolved inconsistency between the methodology and the numerical results, tables, and figures or conclusion of the study. As such, our main studies in question compromised only when there was documentary evidence of retraction; concern, such as a published expression; duplicate publication; unverifiable bibliographic information; materially incomplete content; or some other identifiable research-integrity problem. Similarly, studies that did not meet our criteria were ruled out just because of weak quality or their findings were not in harmony with other studies.

The logic that was adapted for this technical review as follows:

1. Problem/population: AI-enabled WSNs in mission-critical environments

2. Interest/intervention: energy optimisation and edge/security-aware AI techniques

3. Context/comparison: surveillance, smart grid, and urban infrastructure applications

4. Outcome: network lifetime, energy consumption, reliability, latency, security, and scalability.

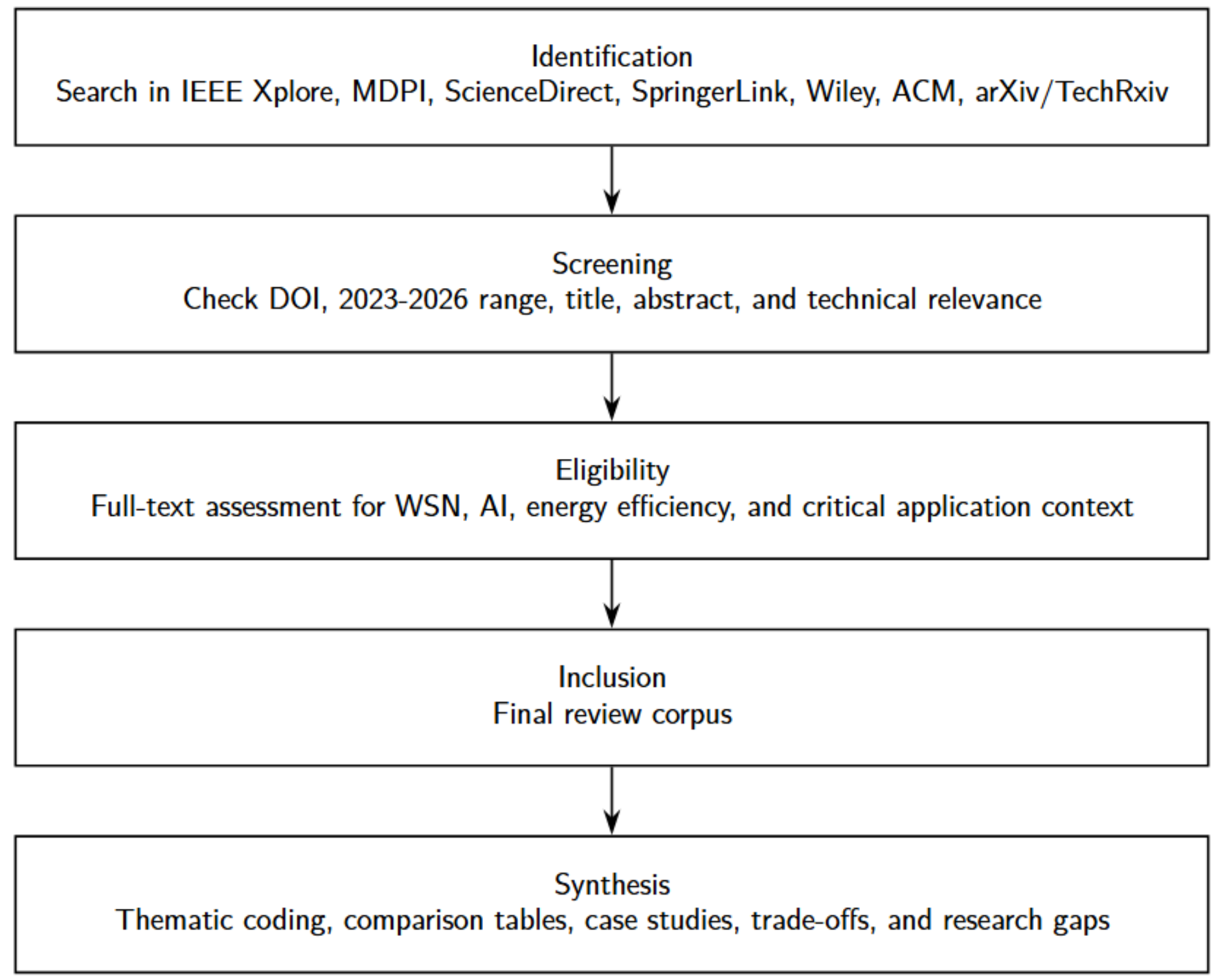

**Figure 2.** PRISMA-style flow diagram for study identification, screening, eligibility assessment, and inclusion.

The thematic classification was finalized after applying the inclusion and exclusion criteria as presented in Figure 2. Only articles published or made available with a DOI between 2023 and 2026, relating to WSNs or WSN-related IoT sensor networks, were included if they addressed at least one of the following dimensions: energy-efficient routing, energy-efficient clustering, edge AI, AI-enabled resource allocation, WSN security, smart grid monitoring, smart city infrastructure, or mission-critical operation. Papers that upon manual inspection and reading were deemed incomprehensible, or whose data was inaccurate or compromised, were also excluded from the review.

Each selected source was coded across four categories:

- Technical approach recorded whether the paper applied routing, clustering, reinforcement learning, fuzzy logic, deep learning, graph neural networks, metaheuristic optimisation, edge AI, or a hybrid method.
- Energy contribution recorded whether the paper aimed to reduce communication cost, extend network lifetime, balance load, reduce control overhead, improve cluster-head selection, or support adaptive resource management.
- Application domain reveals if the work focused on a surveillance/mission-critical system, smart grid, urban infrastructure or a general WSN design transferrable to a critical system.
- Metrics category consisted of network lifetime, delay, throughput, packet delivery ratio, residual energy; number of alive nodes, coverage, scalability; and security detection accuracy.

This coding scheme made it possible to compare dissimilar studies. During extraction, we also logged at the study level any information, where available, on the specific AI or optimisation algorithm, experimental or simulation environment, dataset or application scenario, network size and node configuration, comparison baseline, energy-related metrics, security-related metrics, and explicitly reported limitations. As such, our results section of the manuscript includes comparative synthesis in thematic classification using these study-specific characteristics. This means that if a publication did not report one of these elements, then that information was treated as not reported, i.e. the authors did not infer or reconstruct. For example, [12,13] offer broad review perspectives on AI-based and cluster-based routing. Other studies, provide contemporary examples of hybrid energy-aware routing and clustering [46,47]. Similarly, [16,29], contribute to the discussion on edge AI and smart city sensing. Authors in [52,53] address WSN cybersecurity in smart grids. Though they address disparate challenges, these works help to find common design issues across domains.

The method deals precisely with the published literature indexed by Doi. Because of operational security concerns, many industrial or other systems and real-world deployment of such systems are not publicly available. Another limitation is the lack of shared metrics across studies. Consequently, the review does not intend to rank the algorithms numerically. Rather, it aims at classes of methods, and the conditions under which one is useful.

As such, the goal of this method is practical usefulness. The objective of the review is to evaluate whether an approach is feasible for monitoring or other imperative situations given that mission-critical networks are never designed for a single task. Organizations need to function with limited capabilities despite faults, attacks, and demands to always be operational. As such, in Table 4 we showcase the literature selection and classification criteria.

**Table 4.** Literature selection and classification criteria.

| Methodological dimension | Inclusion criterion | Exclusion criterion | Purpose in the review |
|---|---|---|---|
| Time span | Publications from 2023-2026 | Sources before 2023 | Focus on recent AI-enabled WSN work |
| Bibliographic validity | Mandatory DOI | Sources without DOI | Traceable and verifiable reference base |
| Technology field | WSNs, IoT-based WSNs, edge sensor networks | General IoT or AI papers without WSN relevance | Keep the review thematically focused |
| Energy dimension | Energy efficiency, routing, clustering, lifetime, low-power operation | Papers without energy or network evaluation | Link each source to the review objective |
| AI dimension | ML, DL, RL, fuzzy logic, metaheuristics, edge AI | Classical protocols without AI component | Focus on AI-enabled WSNs |
| Application domain | Monitoring, smart grid, urban infrastructure, critical systems | Unrelated consumer applications | Match the mission-critical scope |

## 5. Results

Classifying the sources produced five main groups of findings. Routing and clustering are among the most frequently recurring concerns in recent AI-enabled WSN research. Energy efficiency means different things depending on whether the setting is monitoring, smart grid, or urban infrastructure. Most solutions trade off energy against security, latency, reliability, and other factors. The literature also lacks shared evaluation benchmarks and enough real-world testing.

The first finding is that routing and clustering remain the strongest, most heavily worked areas. This makes sense, since communication is usually the most energy-costly thing a WSN node does. A bad routing path, or the same nodes used repeatedly as relays, can kill nodes early and break the network. AI-driven routing tries to avoid this by weighing factors beyond what classical protocols use, such as residual energy, distance, link quality, congestion, and node role [12], [13]. Recent work leans toward hybrid models: use a hybrid adaptive optimisation strategy [46], while others combine PSO, fuzzy logic, and self-organising maps [47]. These hybrids matter because efficiency is rarely governed by just one variable.

The second finding is that edge AI is becoming an important part of the architecture. Processing data locally lets a WSN cut down on raw data transmission and react to events faster at the source [16,29]. This matters across all three domains: it lowers transmission cost and helps real-time applications generally, speeds up fault or attack detection in smart grids, and cuts reliance on remote processing in monitoring settings. But edge AI is not free, it can reduce radio energy use while raising local computing load. So the best design isn't the one that uses the most AI, but the one that uses AI selectively, where it actually pays off.

The third finding concerns how application domains differ. All tactical settings need autonomy, secure routing, network resilience, and dependable operation under uncertainty. Smart grids need accuracy, availability, cybersecurity, and tight integration with energy management functions [52,53], [55]. Urban infrastructure needs scalability, low cost, dense coverage, and long-term maintainability [30,31], [66]. Because of these differences, a protocol built for air quality monitoring won't necessarily work for battlefield surveillance or power-line monitoring. This supports designing with the domain in mind from the start.

The fourth finding concerns security. In mission-critical systems, security can't be bolted on afterward. Eavesdropping, selective forwarding, spoofing, and injected malicious routing data can derail a mission even when the network looks energy-efficient on paper [52,53], [56], [58]. Trust-aware and attack-aware routing are therefore essential — though security itself costs energy, which means energy and security need to be optimised together.

The fifth finding is that evaluation methods across studies don't line up. Studies differ in network size, starting energy levels, simulation tools, timing metrics, and baseline protocols. Common metrics - network lifetime, residual energy, packet delivery ratio, throughput, delay, number of alive nodes - often aren't reported together. Security papers may focus on detection accuracy while ignoring energy cost, while energy papers may report long lifetimes without saying how the system holds up under attack. This makes direct comparison across studies hard, which limits how useful many individual results really are in practice.

Taken together, these findings suggest AI-enabled WSNs have real potential for mission-critical environments, but the field isn't equally mature across the board. Energy-efficient routing and clustering are the most technically developed areas. Security work tied to smart grids is growing fast. Urban infrastructure applications are comparatively closer to real-world deployment. Process monitoring-oriented work tends to show up indirectly, more often under terms like secure routing or extreme environments than under explicit monitoring framing. The next step for the field is testing these systems together, under more realistic operating conditions. In Table 5 we provide a categorization of various AI-enabled techniques based on their energy contribution in WSNs. Throughout the paper, the most common technical topic was routing and clustering whereas edge AI, reinforcement learning, fuzzy logic, metaheuristic optimisation and AI-based security were some complementary or more specific ones. Furtheremore, various combinations of the metrics like energy, reliability, latency, packet-delivery, security, coverage, scalability etc. were used for monitoring and mission-critical systems, smart grids, urban infrastructure, and general WSN architectures. As a result, because many articles investigated hybrid methods, various metrics, or more than one application context, they were not treated as mutually exclusive, nor did they yield a strict numerical ranking. This means that the relevant wording is therefore revised so that routing and clustering are described as the most frequently recurring themes in the literature reviewed rather than quantitatively dominant categories. Moreover, in Table 6 we present a comparative table of the needed requirements per application domain. Lastly, Table 7 showcases the main trade-offs in AI-enabled energy-efficient WSNs whereas Table 8 focuses on the research questions posed and Table 9 to the gaps identified.

**Table 5.** Main AI-enabled techniques and their energy contribution in WSNs.

| Technique category | Representative sources | Main energy contribution | Main limitation |
|---|---|---|---|
| AI-driven routing | [12], [46], [50], [65] | Adaptive path selection and reduction of inefficient transmissions | Requires good network-state estimation |
| Cluster-based optimization | [13], [47], [64], [71], [72], | Load balancing and longer node lifetime | Re-clustering overhead |
| Fuzzy logic | [13], [47], [73], [74] | Decision-making under uncertain or multi-criteria conditions | Rules may not transfer across deployments |
| Metaheuristic / bio-inspired methods | [46], [68], [75], [76], [77], [78] | Optimization of cluster heads, paths, and energy distribution | Possible computational overhead |
| Reinforcement learning | [50], [65] | Learning better routing decisions over time | Training cost and reward design |
| Edge AI | [29], [16], [65] | Lower traffic and lower latency through local processing | Computation moves to gateways or edge nodes |
| AI-based security | [52,53], [56-58], [79] | Attack detection and safer routing | Additional energy cost |

**Table 6.** Comparison of requirements by application domain.

| Application domain | Main requirements | Critical AI/WSN functions | Sources |
|---|---|---|---|
| Monitoring / mission-critical systems | Autonomy, security, resilience, hostile operation | Secure routing, trust-aware routing, green routing, anomaly detection | [57,58], [67], [79] |
| Smart grid | Reliability, cybersecurity, accurate monitoring, real-time operation | Attack detection, SWIPT-aware sensing, edge monitoring, energy-aware IoT | [52-55], [65] |
| Urban infrastructure | Scalability, low cost, dense coverage, long-term operation | Edge AI, pollution monitoring, smart lighting, adaptive urban sensing | [29-31], [66] |
| Industrial/extreme cyber-physical systems | Robustness, green routing, adaptive response | AI-based cluster routing and resilient optimization | [67], [77], [80] |

**Table 7.** Main trade-offs in AI-enabled energy-efficient WSNs.

| **Trade-off** | **Description** | **Importance for mission-critical settings** |
|---|---|---|
| Energy vs. security | Attack detection and trust management consume resources | Security cannot be removed in monitoring or smart grid systems |
| Energy vs. latency | Reducing transmissions or duty cycling may delay critical data | Important for surveillance, grid faults, and alerts |
| Energy vs. accuracy | Lower sampling saves energy but may reduce detail | Important in pollution and power-line monitoring |
| Edge AI vs. low-power nodes | Edge reduces traffic but requires computation | Tasks must be distributed across node, gateway, and cloud |
| Clustering vs. stability | Re-clustering balances energy but adds overhead | Problematic in mobile or tactical networks |
| Optimization vs. interpretability | Complex models may perform well but are hard to inspect | Important in critical infrastructure decisions |

**Table 8.** Research questions and corresponding findings.

| **Research question** | **Main finding** | **Sources** |
|---|---|---|
| Which AI techniques dominate? | Routing, clustering, fuzzy logic, metaheuristics, RL, and edge AI | [12,13], [46,47], [50], [65] |
| How is energy efficiency improved? | Load balancing, fewer transmissions, better cluster-head selection, adaptive routing | [46,47], [68], [71], [72], [64] |
| How do application domains differ? | Surveillance and Monitoring: security/resilience; smart grid: | [29], [31], [52,53], [66,67] |
| | reliability/cybersecurity; urban: scale/cost | |
| What is the role of edge AI? | It reduces latency and traffic but moves cost to edge processing | [16], [29], [65] |
| What are the main gaps? | Limited benchmarks, simulation dependence, weak energy-security integration | [12], [13], [53], [78] |

**Table 9.** Main research gaps identified in the review.

| Research gap | Description | Possible direction |
|---|---|---|
| Lack of common benchmarks | Different scenarios, node counts, models, and metrics | Shared testbeds for AI-enabled WSNs |
| Limited field validation | Many results are simulation-based | Field trials in smart grid, urban, and tactical settings |
| Energy and security studied separately | Energy work often ignores attack resilience | Joint energy-security optimization |
| AI computational cost | AI models may increase node or gateway consumption | Lightweight AI and edge/cloud task partitioning |
| Limited interoperability | Different protocols and platforms reduce integration | Standardized architectures and APIs |
| Low interpretability | Complex AI models are hard to inspect | Explainable AI for critical infrastructure |

## 6. Discussion

The study shows that energy metrics should not be the sole concern of AI-enabled WSNs. Low power consumption in a mission-critical environment is only useful when the network remains reliable, safe, and responsive. A network that reduces message transmission to save energy may miss important events. Similarly, a routing scheme that selects the least-cost path may still be insecure if one of the nodes is compromised. The design challenge, then, is not only reducing energy consumption, but balancing energy against operational needs.

Routing illustrates this balance clearly. AI-driven routing offers better performance than existing protocols by incorporating more information and adapting more flexibly [12,13]. It is particularly useful when nodes fail, topology changes, or traffic becomes uneven. However, more sophisticated decision-making requires computation, and may involve more control messages to transmit that information. The energy saved through better routing must therefore exceed the energy spent running the decision mechanism itself — a point that simulation-based studies sometimes fail to examine closely enough.

Clustering faces similar trade-off Cluster-based designs enable local data aggregation and minimizes transmissions to the sink [13], [47], [64], [71]. This is especially worthwhile in networks with a large size and stable topology. Nevertheless, this approach isn't always the best. In mobile, hostile and unstable networks, frequent re-clustering can lead to excessive overheads and instability of service. Clustering is well-suited for smart grids with many fixed monitoring points, but tactical scenarios require greater fault tolerance and faster restoration.

Edge AI has much promise but needs to be treated with caution. The reduction in unnecessary traffic and latency caused by transmission of only relevant data can lead to improved functioning of the overall system. This benefit can also be applied to smart cities, smart grids and moniotirng systems. Although Edge AI comes with its own expenses; nevertheless, it moves some of the energy load from transmission to computing. This trade-off is a win-win when the edge device has enough resources, but it loses its luster when AI inference runs on very constrained nodes. As per the requirements, futuristic architectures must dynamically assign tasks to sensor nodes, cluster heads, gateways, edge nodes, and clouds. As such, a typical WSN-IoT architecture with sensing, multi-hop communication, gateway/edge transport, and application layers is presented in Figure 3.

Safety is also another factor. In routine surveillance systems, data loss can be tolerated, however, in smart grids and monitoring systems, it can't be. A WSN which can easily be attacked is worthless in operation even if it is energy efficient. The literature on eavesdropping detection, selective forwarding detection, secure routing, and trust-based routing highlights the increasing importance of security in WSN design efforts [52,53], [56-58], [79]. Even so, these mechanisms' energy cost must be carefully assessed. If a security layer drains nodes too quickly, it jeopardizes mission reliability.

The three application domains represent different priorities. Networks that focus on monitoring and surveillance of processes and operations must be independent, robust, and secure and work without conventional maintenance. Smart grids necessitate precision, readiness, attack immunity, and cyber-physical stability. Infrastructure development in urban areas suffers limitations by scale, cost, interoperability, and maintainability. Due to mission profile, there is no single best protocol; there are various design choices.

The findings suggest a need for improved evaluation practices. The simulation test alone does not justify the use case for mission-critical applications, but researchers still do it. Real networks face interference, which leads to hardware variations, environmental change and physical damage due to mobility and attack, which cannot be fully tested without real test. Algorithms should be pitted against benchmarks that are commonly used.

Interpretability also matters. Depending on their complexity, some AI techniques behave as black boxes, which creates problems for critical infrastructure. It is essential that engineers know why a path was chosen, why a node was isolated and why some behaviour was flagged as malicious. As a result, explainable AI is not just a research nicety but is core to trust and accountability in these systems.

In summary, AI can improve WSN energy efficiency when calibrated against overall system design. Future systems will likely combine lightweight AI, edge processing, secure routing, adaptive clustering, and interpretable decision-making. The research field should shift focus from isolated algorithmic improvements toward robust, complete architectures tested and validated in real mission-critical environments.

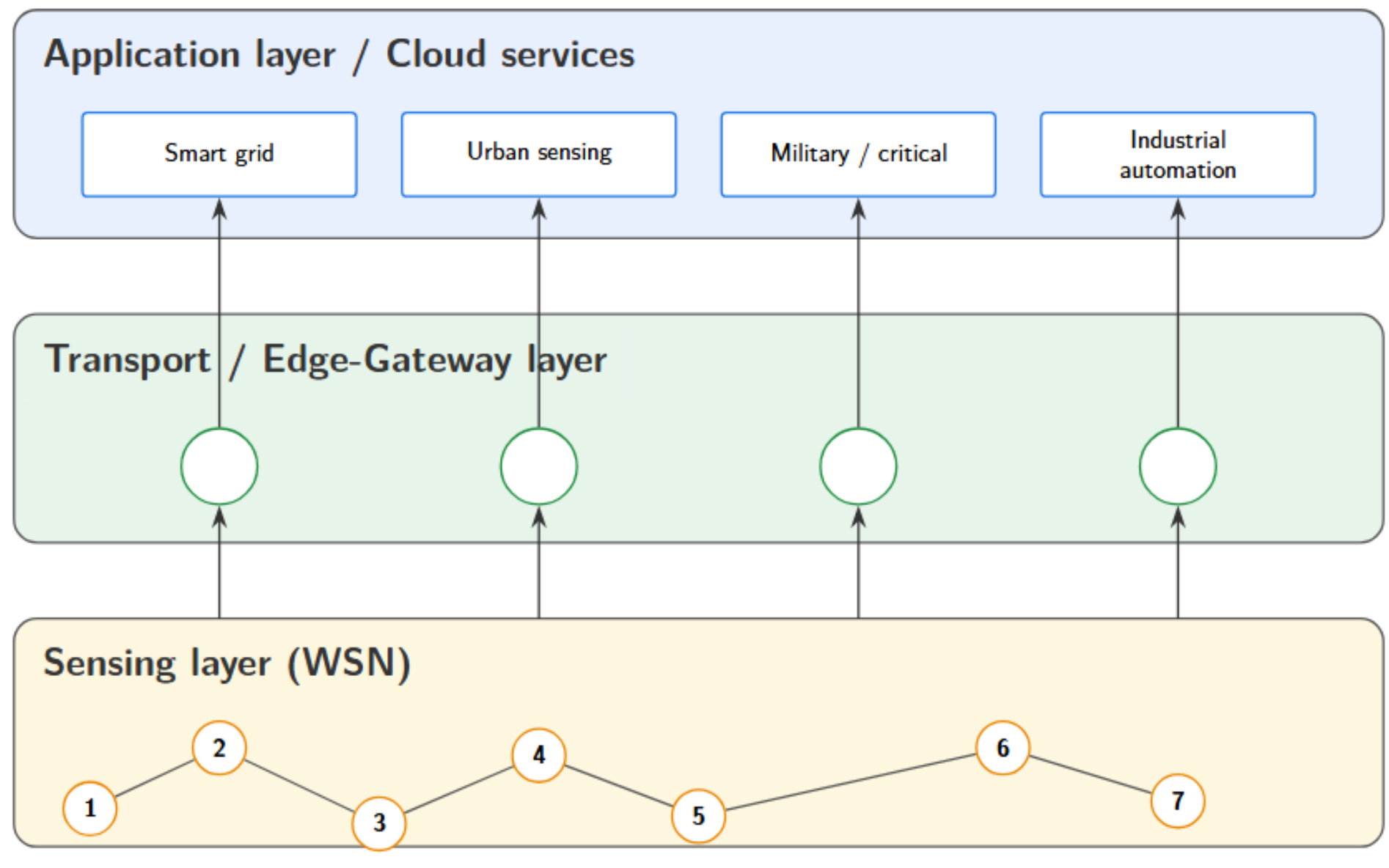

**Figure 3.** Typical WSN-IoT architecture with sensing, multi-hop communication, gateway/edge transport, and application layers

## 7. Conclusions

This review evaluated recent literature on energy-efficient AI-enabled WSNs suitable for mission-critical environments, focusing on monitoring, smart grid, and urban infrastructure applications. The bottom line is that AI-enabled WSNs are not simply enhanced versions of classical sensor networks - they are adaptive cyber-physical sensing systems capable of making autonomous decisions based on energy, topology, traffic, security, and application requirements [12,13], [16]. Clustering algorithms in particular stand out for critical systems, given the important functionality they provide.

Routing and clustering remain the central problems to be solved. Energy efficiency is achieved primarily by minimising redundant transmissions, balancing node load, improving cluster-head selection, and avoiding repeated use of the same paths, which would otherwise deplete specific nodes. AI enhances these processes by introducing adaptiveness and multi-criteria decision-making. However, more complex algorithms also add overhead - the most advanced model is not always the best choice. What matters more is the model that achieves the best balance between energy savings and implementation cost.

The second conclusion is that context of use matters significantly. Surveillance of operations of generic monitoring deployments and infrastructure require autonomy, security, resilience, and tolerance of hostile conditions. Smart grids require accuracy, availability, cybersecurity, and reliable integration with energy management systems (EMSs) [52-55], [65]. Urban infrastructure must be low-cost, scalable, interoperable, and easy to maintain [29-31], [66]. A protocol or architecture does not exist in isolation, so its evaluation must be carried out against the specific mission it is intended to support.

Edge AI is a significant future direction. It can reduce communication load and latency while supporting local detection of important events [16,29]. However, edge AI should not be treated as an automatic energy-saving solution, it shifts where energy is consumed rather than eliminating that consumption. Future designs should dynamically decide where each operation should run: at the sensor node, cluster head, gateway, edge server, or cloud.

The fourth conclusion is that energy and security need to be optimised jointly. In critical systems, a network that is energy-efficient but insecure is not acceptable. At the same time, if security mechanisms consume excessive energy, network lifetime suffers. Future studies should therefore focus on joint energy-security optimisation, particularly for smart grid and monitoring-style applications [52,53], [56], [58], [79].

Several directions exist for future research. First, lightweight AI and TinyML models are needed for low-power sensor nodes. Second, edge/cloud task partitioning should become adaptive, responding dynamically to current energy, delay, and security conditions. In addition, common benchmarks and diversity testbed are needed so algorithms can be tested in similar environments and performance on similar tasks can be compared fairly and effectively; This is particularly important for algorithms making decisions the outcomes of which affect critical infrastructure. In the end, smart grid monitoring, urban sensing, and tactical or extreme environments need more real-world trials.

All in all, AI-enabled WSNs improve energy efficiency only when developed as a part of a complete system. These elements must be collectively considered for optimum results. Studies in this domain should focus towards developing secure, interpretable, lightweight WSN architectures that are tested in the field for real mission-critical applications, rather than optimizing individual protocols.

**Author Contributions:** All authors contributed equally in all stages of this article.

**Funding:** This research received no external funding.

**Data Availability Statement:** Not applicable.

**Acknowledgments:** None.

**Conflicts of Interest:** The authors declare no conflicts of interest.